\documentclass[
twocolumn,
prc,
amsmath,
amssymb,
superscriptaddress
]{revtex4-1}

\usepackage{graphicx}% Include figure files
\usepackage{dcolumn}% Align table columns on decimal point
\usepackage{bm}% bold math
\usepackage{subfigure}
\usepackage{hyperref}
\usepackage{amsmath}
\usepackage{hyphenat}
\usepackage{amsfonts}
\usepackage{booktabs}
\usepackage{multirow}
\usepackage{textcomp}
\hypersetup{bookmarks=true, colorlinks=true,linktoc=page, linkcolor=blue, citecolor=blue, urlcolor=blue }
\graphicspath{%
}%

\begin{document}

\preprint{APS/123-QED}

\title{Systematic study of $\Delta$(1232) resonance excitations using single isobaric charge-exchange reactions induced by medium-mass projectiles of Sn
}
\author{J.~L.~Rodr\'{i}guez-S\'{a}nchez}
\email[Corresponding author:]{joseluis.rodriguez.sanchez@usc.es}
\affiliation{IGFAE, Universidad de Santiago de Compostela, E-15782 Santiago de Compostela, Spain}
\affiliation{GSI-Helmholtzzentrum f\"{u}r Schwerionenforschung GmbH, D-64291 Darmstadt, Germany}
\author{J.~Benlliure}
\affiliation{IGFAE, Universidad de Santiago de Compostela, E-15782 Santiago de Compostela, Spain}
\author{I.~Vida\~{n}a}
\affiliation{INFN, Sezione di Catania, Dipartimento di Fisica ``Ettore Majorana", Universit\`a di Catania, I-95123 Catania, Italy}
\author{H.~Lenske}
\affiliation{Institut f\"{u}r Theoretische Physik der Justus-Liebig Universit\"{a}t Giessen, D-35392 Giessen, Germany}
\author{J.~Vargas}
\altaffiliation[Present address: ]{Universisdad Santo Tom\'{a}s, 15001 Tunja, Colombia.}
\affiliation{Universidad de Santiago de Compostela, E-15782 Santiago de Compostela, Spain}
\author{C.~Scheidenberger}
\affiliation{GSI-Helmholtzzentrum f\"{u}r Schwerionenforschung GmbH, D-64291 Darmstadt, Germany}
\author{H.~Alvarez-Pol}
\affiliation{IGFAE, Universidad de Santiago de Compostela, E-15782 Santiago de Compostela, Spain}
\author{J.~Atkinson}
\affiliation{GSI-Helmholtzzentrum f\"{u}r Schwerionenforschung GmbH, D-64291 Darmstadt, Germany}
\author{T.~Aumann}
\affiliation{GSI-Helmholtzzentrum f\"{u}r Schwerionenforschung GmbH, D-64291 Darmstadt, Germany}
\affiliation{Institut f\"{u}r Kernphysik, Technische University\"{a}t Darmstadt, D-64289 Darmstadt, Germany}
\author{Y.~Ayyad}
\affiliation{IGFAE, Universidad de Santiago de Compostela, E-15782 Santiago de Compostela, Spain}
\author{S.~Beceiro-Novo}
\altaffiliation[Present address: ]{Department of Physics and Astronomy, Michigan State University, East Lansing,  MI 48824-1321, USA.}
\affiliation{Universidad de Santiago de Compostela, E-15782 Santiago de Compostela, Spain}
\author{K.~Boretzky}
\affiliation{GSI-Helmholtzzentrum f\"{u}r Schwerionenforschung GmbH, D-64291 Darmstadt, Germany}
\author{M.~Caama\~{n}o}
\affiliation{IGFAE, Universidad de Santiago de Compostela, E-15782 Santiago de Compostela, Spain}
\author{E.~Casarejos}
\affiliation{CINTECX, Universidad de Vigo, E-36200 Vigo, Spain}
\author{D.~Cortina-Gil}
\affiliation{IGFAE, Universidad de Santiago de Compostela, E-15782 Santiago de Compostela, Spain}
\author{P.~D\'{i}az~Fern\'{a}ndez}
\altaffiliation[Present address: ]{Institutionen f\"{o}r Fysik, Chalmers Tekniska H\"{o}gskola, 41296 G\"{o}teborg, Sweden.}
\affiliation{Universidad de Santiago de Compostela, E-15782 Santiago de Compostela, Spain}
\author{A.~Estrade}
\altaffiliation[Present address: ]{Department of Physics, Central Michigan University, Mount Pleasant, MI 48858, USA.}
\affiliation{GSI-Helmholtzzentrum f\"{u}r Schwerionenforschung GmbH, D-64291 Darmstadt, Germany}
\affiliation{Saint Mary's University, Halifax, Nova Scotia B3H 3C3, Canada}
\author{H.~Geissel}
\affiliation{GSI-Helmholtzzentrum f\"{u}r Schwerionenforschung GmbH, D-64291 Darmstadt, Germany}
\author{E.~Haettner}
\affiliation{GSI-Helmholtzzentrum f\"{u}r Schwerionenforschung GmbH, D-64291 Darmstadt, Germany}
\author{A.~Keli\'{c}-Heil}
\affiliation{GSI-Helmholtzzentrum f\"{u}r Schwerionenforschung GmbH, D-64291 Darmstadt, Germany}
\author{Yu.~A.~Litvinov}
\affiliation{GSI-Helmholtzzentrum f\"{u}r Schwerionenforschung GmbH, D-64291 Darmstadt, Germany}
\author{C.~Paradela}
\altaffiliation[Present address: ]{EC-JRC, Institute for Reference Materials and Measurements, Retieseweg 111, B-2440 Geel, Belgium.}
\author{D.~P\'{e}rez-Loureiro}
\altaffiliation[Present address: ]{National Superconducting Cyclotron Laboratory, Michigan State University, East Lansing, MI 48824-1321, USA.}
\affiliation{Universidad de Santiago de Compostela, E-15782 Santiago de Compostela, Spain}
\author{S.~Pietri}
\affiliation{GSI-Helmholtzzentrum f\"{u}r Schwerionenforschung GmbH, D-64291 Darmstadt, Germany}
\author{A.~Prochazka}
\altaffiliation[Present address: ]{Niigata University, 8050 Ikarashi 2-no-cho, Nishi-ku, 950-2181 Niigata, Japan.}
\affiliation{GSI-Helmholtzzentrum f\"{u}r Schwerionenforschung GmbH, D-64291 Darmstadt, Germany}
\author{M.~Takechi}
\altaffiliation[Present address: ]{Niigata University, 8050 Ikarashi 2-no-cho, Nishi-ku, 950-2181 Niigata, Japan.}
\affiliation{GSI-Helmholtzzentrum f\"{u}r Schwerionenforschung GmbH, D-64291 Darmstadt, Germany}
\author{Y.~K.~Tanaka}
\affiliation{GSI-Helmholtzzentrum f\"{u}r Schwerionenforschung GmbH, D-64291 Darmstadt, Germany}
\affiliation{High Energy Nuclear Physics Laboratory, RIKEN, Saitama 351-0198, Japan}
\author{H.~Weick}
\affiliation{GSI-Helmholtzzentrum f\"{u}r Schwerionenforschung GmbH, D-64291 Darmstadt, Germany}
\author{J.~S.~Winfield}
%\footnotetext{Deceased.}
%\altaffiliation[]{Deceased.}
\affiliation{GSI-Helmholtzzentrum f\"{u}r Schwerionenforschung GmbH, D-64291 Darmstadt, Germany}

\date{\today}% It is always \today, today,
             
\begin{abstract}
The fragment separator FRS has been for the first time used to measure the $(n, p)$ and $(p, n)$-type isobaric charge-exchange cross sections of stable $^{112,124}$Sn isotopes accelerated at 1$A$ GeV with an uncertainty of $3\%$ and to separate quasi-elastic and inelastic components in the missing-energy spectra of the ejectiles\footnote{This work comprises part of the Ph.D. thesis of J. Vargas.}. The inelastic contribution can be  associated to the excitation of isobar $\Delta$(1232) resonances and to the pion emission in s-wave, both in the target and projectile nuclei, while the quasi-elastic contribution is associated to the nuclear spin-isospin response of nucleon-hole excitations. The data lead to interesting results where we observe a clear quenching of the quasi-elastic component and their comparisons to theoretical calculations demonstrate that the baryonic resonances can be excited in the target and projectile nuclei. To go further in this investigation, we propose to study the excitation of baryonic resonances taking advantage of the combination of high-resolving power magnetic spectrometers with the WASA calorimeter. These new measurements will allow us to determine the momenta of the ejectiles and pions emitted in coincidence after the single isobaric charge-exchange collisions, providing us unique opportunities to study the evolution of the baryonic resonance dynamics with the neutron-proton asymmetry through the use of exotic radioactive ion beams. 

\end{abstract}
\maketitle
\section{Introduction}

The understanding of the nucleon and its excited internal structure, forming the so-called baryonic resonances, still remains a fundamental challenge in hadronic physics. The experimental and theoretical study of the baryon excited states helps to improve our knowledge of Quantum Chromodynamics (QCD)~\cite{Greenberg1964} in the non-perturbative regime, especially for understanding the composition of a baryon as a superposition of hadronic and quarkonic components. Experimental information about the mass, width, and decay modes of baryonic resonances in free space and in nuclear environments is crucial for validating theoretical models describing the internal structure of the nucleon and its excited states. Moreover, the investigations of baryon resonances in nuclear matter are a natural extension of nuclear physics at the intersection of nucleonic and subnuclear degrees of freedom, elucidating open problems as the role of resonances in three-body interactions~\cite{Krebs2018,Feng2016} and neutron star matter~\cite{Li2018,Li2019}.

In the literature, the lowest-lying baryon resonance is the $\Delta$(1232) state. This $\Delta$-resonance is the first spin-isospin excited mode of the nucleon corresponding to a $\Delta S = 1$; $\Delta I = 1$ transition, seen prominently as an elastic p-wave resonance in pion-nucleon scattering. That state is closely connected with the excitation of the well known GT resonance as already seen in nuclear collisions~\cite{Sakai1999,Liang2018}. Charge-exchange excitations of the nuclear GT-states and the nucleonic $\Delta$-resonance are both related to the action of the same spin-isospin transition operators~\cite{Oset1979,Arima2001}. Actually, a long standing problem of nuclear structure physics is to understand the coupling of these two modes.  Here, we would like to remark that in the last decade a large effort has been carried out to understand the quenching problem of nuclear GT-states at low momentum transfer, leading to the development of state-of-the art model calculations~\cite{Gysbers2019} that provide a reasonable description of the experimental data. However, at high momentum transfer the quenching of the GT strength~\cite{Sakai2004} remains still unsolved, which is mainly attributed to a large redistribution of transition strength due to the coupling of $\Delta$-particle--nucleon-hole and purely nucleonic particle--hole excitations~\cite{Lenske2018,Lenske2019}. Similar mechanisms, although not that clearly seen, are present in the high-energy flank of the Fermi-type spectral distribution, i.e. the non-spin flip charge-exchange excitations mediated by the isospin transition operator alone, caused by the coupling to the Roper resonance~\cite{Morsch1992}. Concerning its properties, modifications of the mass and width of the $\Delta$-resonance have been observed in central heavy-ion collisions leading to dense and heated hadron matter. In Ref.~\cite{Hjort1997}, for example, the mass shift and the width were determined as functions of the centrality, both showing a substantial reduction with decreasing impact parameter. The modifications of the $\Delta$ properties has been interpreted in terms of the values of hadronic density, temperature, and various non-nucleon degrees of freedom in nuclear matter~\cite{Hennino1982}. However, many experimental findings on baryon resonance production are obtained after non-trivial analysis that usually result in ambiguities with their theoretical interpretations. Partly, the persisting uncertainties are related to difficulties in extracting the spectral structures from data. As discussed in Ref.~\cite{Trzaska1991}, in heavy-ion collisions a major obstacle for the reconstruction of the resonance spectral distributions is the large background of uncorrelated p$\pi$ pairs coming from other sources. This problem has been found to be less severe in peripheral reactions.

A renewed interest on the in-medium properties of baryon resonances is motivated by the challenge to understand the composition of neutron stars. The use of realistic equations of state with updated values for the density derivative of the symmetry energy yields a threshold density for the appearance of $\Delta$-isobars similar to the one for hyperons~\cite{Drago14,Cai15,Drago16}. Recently, the study of magnetic field effects on neutron star matter has pointed out that the vector-isovector self-interaction for the mesons in combination with a strong magnetic field enhances the $\Delta$-baryon population in dense matter, while decreasing the relative density of hyperons~\cite{Dexheimer2021}. These findings also allow for a better agreement with the observations of neutron star masses, indicating that likely magnetic neutron stars also contain $\Delta$-baryons in their interior.

Pioneering experiments showing for the first time the in-medium excitation of the $\Delta$-resonance in peripheral single isobaric charge-exchange reactions were performed in the 1980’s at the Laboratoire National SATURNE in Saclay (France). In these experiments, the ejectiles produced in collisions of stable relativistic projectiles of $^{3}$He, $^{12}$C, $^{20}$Ne, and $^{40}$Ar on different target materials were identified and momentum analyzed with the magnetic spectrometer SPES-IV~\cite{Grorud1981,Ellegaard1985,Contardo1986,Roy1988}. The high-resolving power of the spectrometer made it possible to prove the excitation of the $\Delta$-resonance from the velocity recoil experienced by the ejectiles due to the pion emission. Slightly later, similar experiments were also perfomed in the Synchro-Phasetron at the JINR Dubna facility as discussed in Ref.~\cite{Lenske2018}.

Following SATURNE's research, the first investigations at the GSI accelerator facility in Darmstadt (Germany) were performed in the 1990's by S\"{u}mmerer and collaborators~\cite{Summerer1995} with medium-mass projectiles of $^{129}$Xe at a kinetic energy of 790$A$ MeV. This experiment was carried out at the fragment separator FRS~\cite{Geissel1992}, in which the individual charge-exchange products were identified by charge and mass number. Their inclusive cross sections were used to investigate the effects due to the $\Delta$-resonance production through the comparison to model calculations, showing that in the $(n, p)$-type isobaric charge-exchange reaction the presence of this resonance can increase the cross section by a factor of two. Latter, more experiments were perfomed by the {\it CHARMS} collaboration employing medium-mass and heavy projectiles of $^{136}$Xe~\cite{Benlliure2008} and $^{209}$Bi~\cite{Kelic2004}, measuring simultaneously the cross sections and missing-energy distributions of the single isobaric charge-exchange products with resolutions of around 50 MeV~\cite{Vargas2013}. More recently, the missing-energy resolution has been improved down to 10 MeV~\cite{Rodriguez2020} by using time-projection chambers (TPCs)~\cite{Janik11} with a position resolution of 300 $\mu m$ full width at half maximum (FWHM) and owing to the reduction of layers of matter in the beam line. 

In the present work, we use single isobaric charge-exchange reactions produced with medium-mass projectiles of $^{112,124}$Sn impinging on different target nuclei, which allow us to cover a large range in neutron excess. The experiment and setup employed for the measurements are described in Sec.~\ref{sec:exp}, while the experimental results are shown in Sec.~\ref{sec:rst}. For the data interpretation we use a theoretical model decribed in Sec.~\ref{sec:model} and the discussion is presented in Sec.~\ref{sec:disc}. Finally, the main conclusions and the future perspectives are given in Sec.~\ref{sec:conclusions}.

\section{Experimental approach}
\label{sec:exp}

\begin{center}
\begin{figure}[h]
\begin{center}
\includegraphics[width=0.47\textwidth,keepaspectratio]{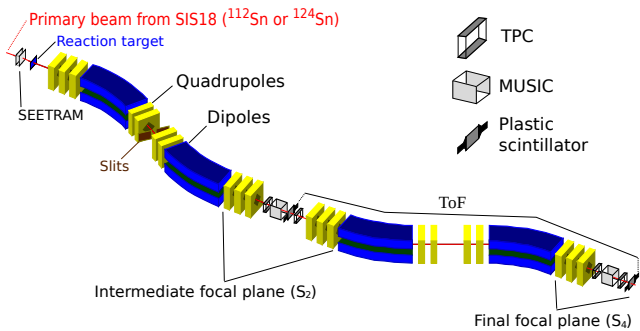}
\caption{(Color online) Schematic view of the FRS experimental setup used in the present work with the reaction target located at the entrance of the FRS spectrometer to induce the single isobaric charge-exchange reactions.
}
\label{fig:1}
\end{center}
\end{figure}
\end{center}

The experiment has been carried out at the GSI facility using the SIS-18 synchrotron combined with the FRS spectrometer~\cite{Geissel1992}. The FRS is a zero-degree high-resolving power spectrometer with a typical resolving power of $B\rho/\Delta B\rho =1500$ and with an angular acceptance of $\pm$15 mrad around the central trajectory. Beams of $^{112}$Sn and $^{124}$Sn, accelerated up to kinetic energies of 1$A$ GeV, were delivered by the SIS-18 with an intensity around 10$^{8}$ ions per spill and guided up to the FRS entrance to produce the single isobaric charge-exchange reactions in different target nuclei. The FRS spectrometer was used in the achromatic mode, where the reaction target is placed at the FRS entrance and the full spectrometer is utilized to separate and to identify the projectile residues, as shown in Fig.~\ref{fig:1}.

The measurements were performed using polyethylene (CH$_{2}$), Carbon, Copper, and Lead targets with thicknesses of 95, 103, 373 and 255 mg$/$cm$^{2}$, respectively. The experimental data with the carbon target was also used to subtract the contribution of the carbon-induced reactions from the CH$_{2}$ target and obtain the proton contribution. An additional measurement was performed without target in order to obtain the background of reactions coming from other layers of matter located at the FRS target area, which was of around $1 \%$.

The magnetic ridigity $B \rho$ of each reaction product passing through the FRS can be obtained with the TPC detectors~\cite{Janik11} located at the FRS focal planes on an event-by-event basis. For the intermediate focal plane (S2) the magnetic ridigity ($B \rho^{S2}$) was determinated as follows:
\begin{equation}\label{eq:planeS2}
 B \rho^{S2} =B \rho_{0}^{ref} \left(1- \frac{x_{2} }{ D_{S2}} \right),
\end{equation}
where $x_{2}$ is the horizontal position at the intermediate focal plane, $D_{S2}$ is the value of the dispersion from the production target up to the focal plane S2, and $B \rho_{0}^{ref}$ is the magnetic ridigity of a central trajectory along the first section of the FRS.

In the case of the final focal plane (S4), the magnetic ridigity ($B \rho^{S4}$) was obtained according to: 
\begin{equation}\label{eq:planeS4}
B \rho^{S4}=B \rho^{ref}_{0} \left(1- \frac{x_{4}-M x_{2} }{ D_{S4}} \right),
\end{equation}
where $x_{4}$ is the position at the final focal plane S4, $D_{S4}$ is the dispersion between both focal planes, S2 and S4, and $M$ is the magnification.

In Eqs. (\ref{eq:planeS2}) and (\ref{eq:planeS4}), the magnetic rigidity of the central trajectory as well as the dispersions and the magnification were determined by measuring the trajectory of $^{112}$Sn projectiles at 1$A$ GeV for different values of the magnetic fields in the dipoles of the spectrometer. These measurements allowed us to calibrate the FRS optics, obtaining for the middle and final focal plane dispersions $D_{S2}=-7.48$ cm/$ \% $ and $D_{S4}=7.56$ cm/$ \% $, respectively, with an uncertainty of $0.02$ cm/$ \% $ in both cases. These quantities also provided us the magnification of the spectrometer according to $M=D_{S4}/D_{S2}$, resulting in $M$=($-1.011 \pm 0.004$).

Nuclei transmitted through the FRS were identified in mass over charge ($A/q$) through the determination of the magnetic ridigity at the final focal plane (see Eq. (\ref{eq:planeS4})) and the fragment velocity ($v$) according to Refs.~\cite{Teresa14,JL2017}:
\begin{equation}
\frac{A}{q} = \frac{ e B \rho^{S4} }{ u \gamma \beta c },\nonumber
\end{equation}
where $q$ is the atomic number if we assume that the fragments were completely stripped ($q=Z$), $u$ is the atomic mass unit, $e$ is the elementary charge, $c$ is the speed of the light, $\beta=v/c$ and $\gamma=1/\sqrt{1-\beta^2}$. The velocity of fragments was obtained by combining the S2-S4 path length with the time-of-flight (ToF) measurements performed with the plastic scintillators located at the focal planes of the FRS, which provided the ToF with a resolution of around 150 ps (FWHM). Additionally, the measurement of the atomic number $Z$ was provided by two multi-sampling ionization chambers (MUSIC) located in the focal planes of the FRS.

\begin{center}
\begin{figure}[h]
\begin{center}
\includegraphics[width=0.48\textwidth,keepaspectratio]{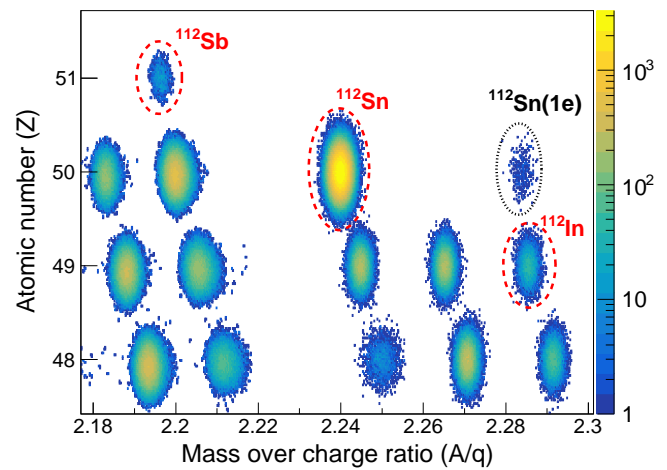}
\caption{(Color online) Identification matrix at the final focal plane of the FRS obtained from reactions induced by $^{112}$Sn impinging on the carbon target where one can identify the single isobaric charge-exchange residues $^{112}$Sb and $^{112}$In. In addition, the single electron charge state of $^{112}$Sn is marked by a dotted ellipse.
}
\label{fig:2}
\end{center}
\end{figure}
\end{center}

As a result, Fig.~\ref{fig:2} displays the identification matrix of the fragmentation residues produced by the projectiles of $^{112}$Sn impinging on the carbon target. The figure was obtained by overlapping three magnetic settings of the FRS centered on the isotopes $^{112}$Sb, $^{112}$Sn, and $^{112}$In. The calibration in atomic and mass number was performed with respect to the signals registered for the beam of $^{112}$Sn. In the figure, one can see the good resolution in atomic and mass number achieved in this measurement: $\Delta Z/Z$=6~$\times$~10$^{-3}$ and $\Delta A/A$=1~$\times$10$^{-3}$ (FWHM), respectively.

\section{Results}
\label{sec:rst}

Total isobaric charge-exchange cross sections were accurately determined by normalizing the production yield of the charge exchange residual nuclei to the number of projectiles and target nuclei. The number of incoming projectiles was obtained by using a secondary electron monitor (SEETRAM)~\cite{Junghans1996} placed at the entrance of the FRS, as shown in Fig.~\ref{fig:1}, which was calibrated with a reference plastic scintillator~\cite{Voss1995}. The uncertainty in the determination of the incoming projectile intensity was of 3$\%$. The number of isobaric charge exchange residues was obtained from the corresponding identification matrix at the final focal plane by gating on the fragment of interest, as shown in Fig.~\ref{fig:2} for the single isobaric charge-exchange residues of $^{112}$Sb and $^{112}$In. The number of counts for these residues was then corrected by the detection efficiency according to the procedure detailed in Refs.~\cite{Teresa14,JL2017}. The first correction factor is related to the use of high-intensity beams, which affect the determination of the cross sections because of the acquisition dead time. This correction was evaluated from the ratio between the number of accepted and total triggers registered by the data acquisition system. This value was kept below 20$\%$ during the experiment by reducing the optical transmission of other fragments. This was carried out by closing the slits located between the first and second FRS dipole (see Fig.~\ref{fig:1}), leaving a hole of 40 mm. The second correction is due to secondary reactions that occur in the different layers of matter along the beam line, which were taken into account by calculating with the KAROL code~\cite{karol75} the total probability of interaction in each layer of matter. The obtained correction factors were less than 2$\%$. The third correction is to take into account the population of atomic charge states due to electromagnetic interactions of the nuclear residues with the different layers of matter along the beam line, which can alter the measurement of the $A/q$ ratio and therefore affects the number of counts used in the determination of the cross section~\cite{Alcantara15,Paradela2017}. In order to determine these losses we used the code GLOBAL~\cite{global}, being this correction less than $1\%$. Finally, we remark that the FRS ion-optical transmission for the isobaric charge-exchange residues investigated in this work was of 100$\%$. The final values obtained for the total isobaric charge-exchange cross sections are listed in Table~\ref{tab:1}.

The missing-energy spectra of the recoiling $(p, n)$ and $(n, p)$-type charge-exchange residues were obtained by measuring the ejectile kinetic energies with respect to the projectile one determined in the middle of the target, being the kinetic energy ($E_{k}$) calculated as:
\begin{equation}
E_{k}= \sqrt{P + M} - M,\nonumber
\end{equation}
where $M$ is the nuclear mass and $P$ is the momentum, obtained as follows:
\begin{equation}
P=B\rho^{S2}Z,\nonumber
\end{equation}
where $Z$ is the atomic number and $B\rho^{S2}$ is determinated according to Eq. (\ref{eq:planeS2}), taking also into account the beam-extraction time correction~\cite{Tanaka2018} to reach the best FRS resolution.

\begin{table}[t]
\caption{Total, quasi-elastic (qe), and inelastic (in) cross sections of the single isobaric charge-exchange reactions measured in the present work. The uncertainties include statistical and systematic contributions.}
\label{tab:1}
\begin{center}
\begin{ruledtabular}
\begin{tabular}{cccc}
Reaction & $\sigma_{Total}$ [$\mu$b] & $\sigma_{qe}$ [$\mu$b] & $\sigma_{in}$ [$\mu$b]  \\
\noalign{\smallskip}\hline\noalign{\smallskip}
p($^{112}$Sn,$^{112}$In)X & 291 $\pm$ 9 & - & 291 $\pm$ 9\\
p($^{112}$Sn,$^{112}$Sb)X & 604 $\pm$ 19 & 235 $\pm$ 10 & 369 $\pm$ 15\\
$^{12}$C($^{112}$Sn,$^{112}$In)X & 705$\pm$ 22 & 223 $\pm$ 9 & 432 $\pm$ 20\\
$^{12}$C($^{112}$Sn,$^{112}$Sb)X & 718 $\pm$ 22 & 276 $\pm$ 16 & 442 $\pm$ 26\\
Cu($^{112}$Sn,$^{112}$In)X & 809 $\pm$ 25 & 307 $\pm$ 18 & 502 $\pm$ 30\\
Cu($^{112}$Sn,$^{112}$Sb)X & 758 $\pm$ 23 & 235 $\pm$ 16 & 523 $\pm$ 33\\
Pb($^{112}$Sn,$^{112}$In)X & 1000 $\pm$ 31 & 465 $\pm$ 28 & 535 $\pm$ 31\\
Pb($^{112}$Sn,$^{112}$Sb)X & 793 $\pm$ 25 & 198 $\pm$ 14 & 595 $\pm$ 39\\
\noalign{\smallskip}\hline\noalign{\smallskip}
p($^{124}$Sn,$^{124}$Sb)X & 585 $\pm$ 19 & 203 $\pm$ 9 & 382 $\pm$ 17\\
C($^{124}$Sn,$^{124}$Sb)X & 705 $\pm$ 22 & 220 $\pm$ 10 & 485 $\pm$ 22
\end{tabular}
\end{ruledtabular}
\end{center}
\end{table}

\begin{figure*}
\begin{center}
\includegraphics[width=0.98\textwidth,keepaspectratio]{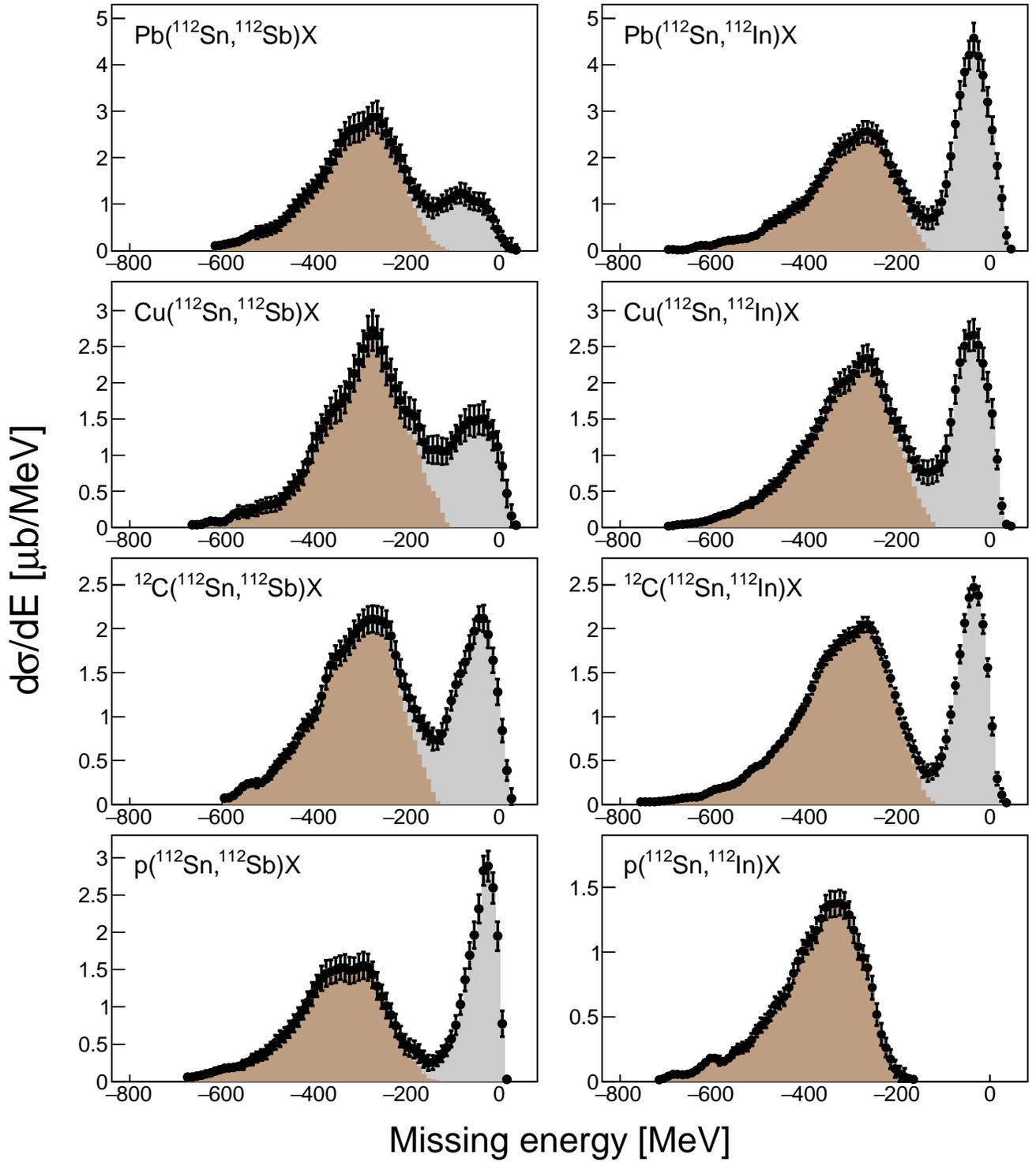}
\caption{(Color online) Missing-energy spectra obtained from the Pb, Cu, $^{12}$C and proton targets for the single isobaric charge-exchange reactions ($^{112}$Sn,$^{112}$Sb) and ($^{112}$Sn,$^{112}$In). The quasi-elastic and inelastic contributions are displayed with gray and brown histograms, respectively.
}
\label{fig:3}
\end{center}
\end{figure*}

\begin{figure}[b]
\centering
\subfigure{
\includegraphics[width=0.48\textwidth,keepaspectratio]{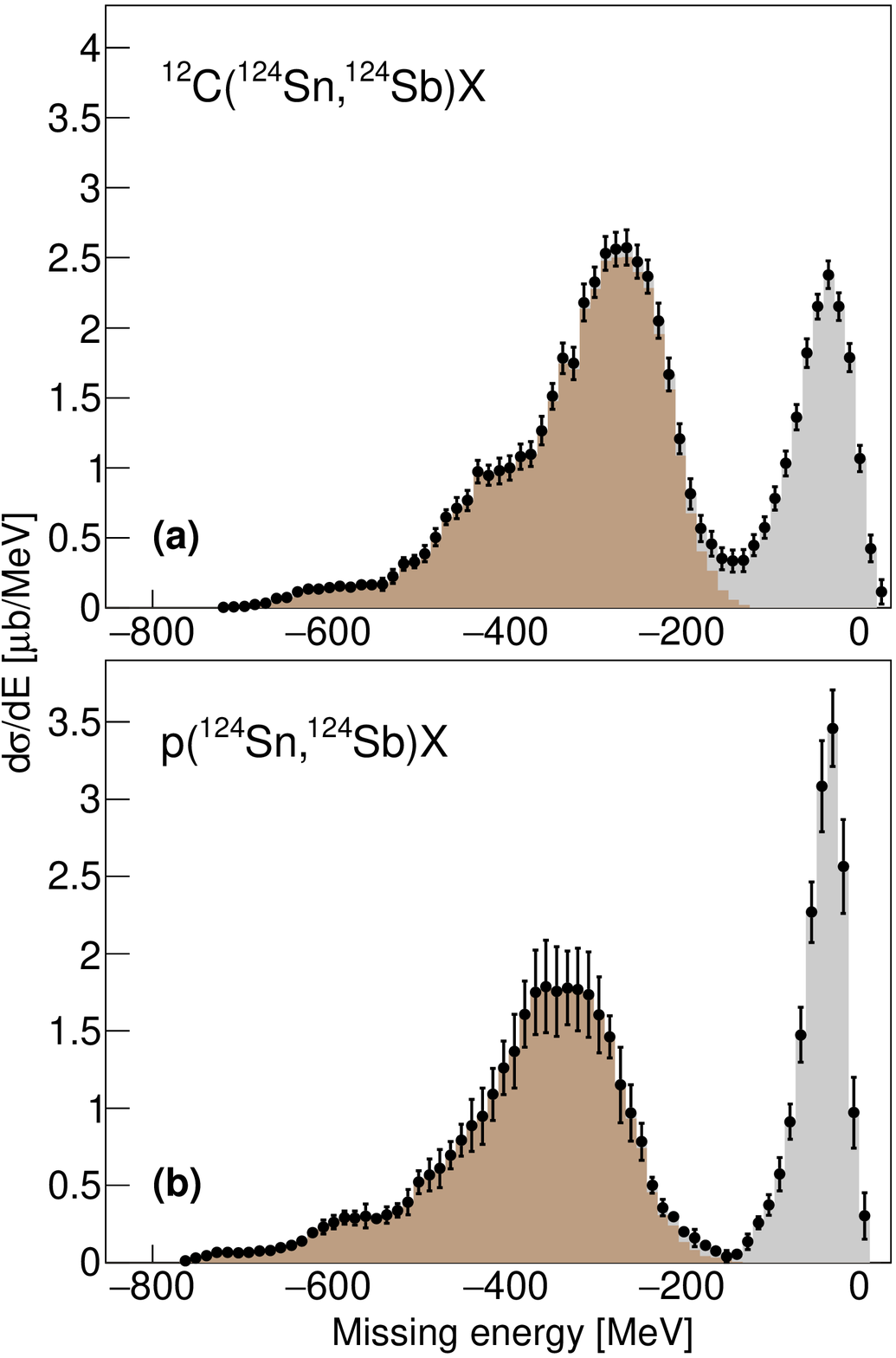}}
\caption{(Color online) Missing-energy spectra obtained from the $^{12}$C (a) and proton (b) targets for the single isobaric charge-exchange reaction ($^{124}$Sn,$^{124}$Sb). The quasi-elastic and inelastic contributions are displayed with gray and brown histograms, respectively.
}
\label{fig:4}
\end{figure}

The results are displayed in Fig.~\ref{fig:3}, where we show the missing-energy spectra obtained from the $(n, p)$ (left panels) and $(p,n)$ (right panels) isobaric charge-exchange reactions induced by ions of $^{112}$Sn in the Pb, Cu, $^{12}$C and proton targets at projectile kinetic energies of 1$A$ GeV, which were normalized to the cross section of each reaction (see Table~\ref{tab:1}). The contribution of the proton target was obtained from the CH$_{2}$ spectrum by subtracting the carbon contribution, according to the equation:
$\sigma_p = 0.5 \times [\sigma_{CH_2} - \sigma_C]$, applied on a bin-by-bin basis. For all the spectra, the background was subtracted on a bin-by-bin basis as well. The same methodology was applied to the $(n, p)$-type isobaric charge-exchange measurements performed with the primary beam of $^{124}$Sn impinging on the $^{12}$C and proton targets, displayed in the upper and lower panels of Fig.~\ref{fig:4}, respectively. Thanks to the high-resolving power of the FRS together with a sizeable reduction of matter along the beam line, we succeed to measure these spectra with a resolution of 10 MeV, which is a factor of two better than the one reached in the SATURNE experiments~\cite{Roy1988}. Because these spectra are affected by the energy and angular straggling of ions passing through the targets, as observed in previous experiments~\cite{Kelic2004}, an unfolding procedure based on the Richardson-Lucy's technique~\cite{Lucy1974} together with a regularization method to optimize the stability of the solution against statistical fluctuations~\cite{Vargas2013} was used to improve the sharpness of the spectra. This unfolding approach was applied to both Sn projectiles using for the FRS response function the missing-energy spectrum of the primary beam passing through the corresponding target after correcting the local energy-straggling due to the exchange of a single proton with the target nucleus.

These spectra show that the spin-isospin response of our projectile nuclei is concentrated in two energy domains, in which the momentum transfer is larger than 100 MeV/c. The first group of excitations starting at missing energies close to zero is of quasi-elastic origin. This region is dominated by quasi-free charge-exchange processes between projectile and target nucleons, primarily reflecting global intrinsic properties of the colliding nuclei like Fermi motion and, to a lesser extent, isovector mean-field dynamics. Collective isovector multipole modes could contribute as well, but their identification requires a multipole decomposition of the spectra as perfomed for light ion reactions~\cite{PhysRevC.86.014304}. In this region, nucleon-nucleon correlations could also contribute to the spectra by increasing the kinetic energy of the outgoing isobaric charge-exchange residues~\cite{Litvinova2016}, resulting in positive missing-energy tails observed to the right of the quasi-elastic peaks. Note that the quasi-elastic contribution for the ($^{112}$Sn,$^{112}$In) reaction with the proton target disappears, as expected, because this transition needs a target containing neutrons. At lower missing energies, the observed bump corresponds to a nucleon being excited into a $\Delta$-resonance in the target or projectile nuclear systems. This twofold spectrum is a common feature of all isobaric charge-exchange reactions induced at high kinetic energies (more than 0.4$A$ GeV) and illustrates clearly that the $\Delta$ is the first spin-isospin excited mode of the nucleon, corresponding to a $\Delta S=1$ and $\Delta I=1$ spin-isospin change as forementioned above~\cite{Lenske2018}. In the spectra shown in Figs.~\ref{fig:3} and~\ref{fig:4}, one can also see a clear energy shift of around 63 MeV for the inelastic peak between the proton and the other targets. This will be discussed later in Sec.~\ref{sec:disc}.

\begin{figure}[h]
\centering
\subfigure{\label{fig:5a}\includegraphics[width=0.48\textwidth,keepaspectratio]{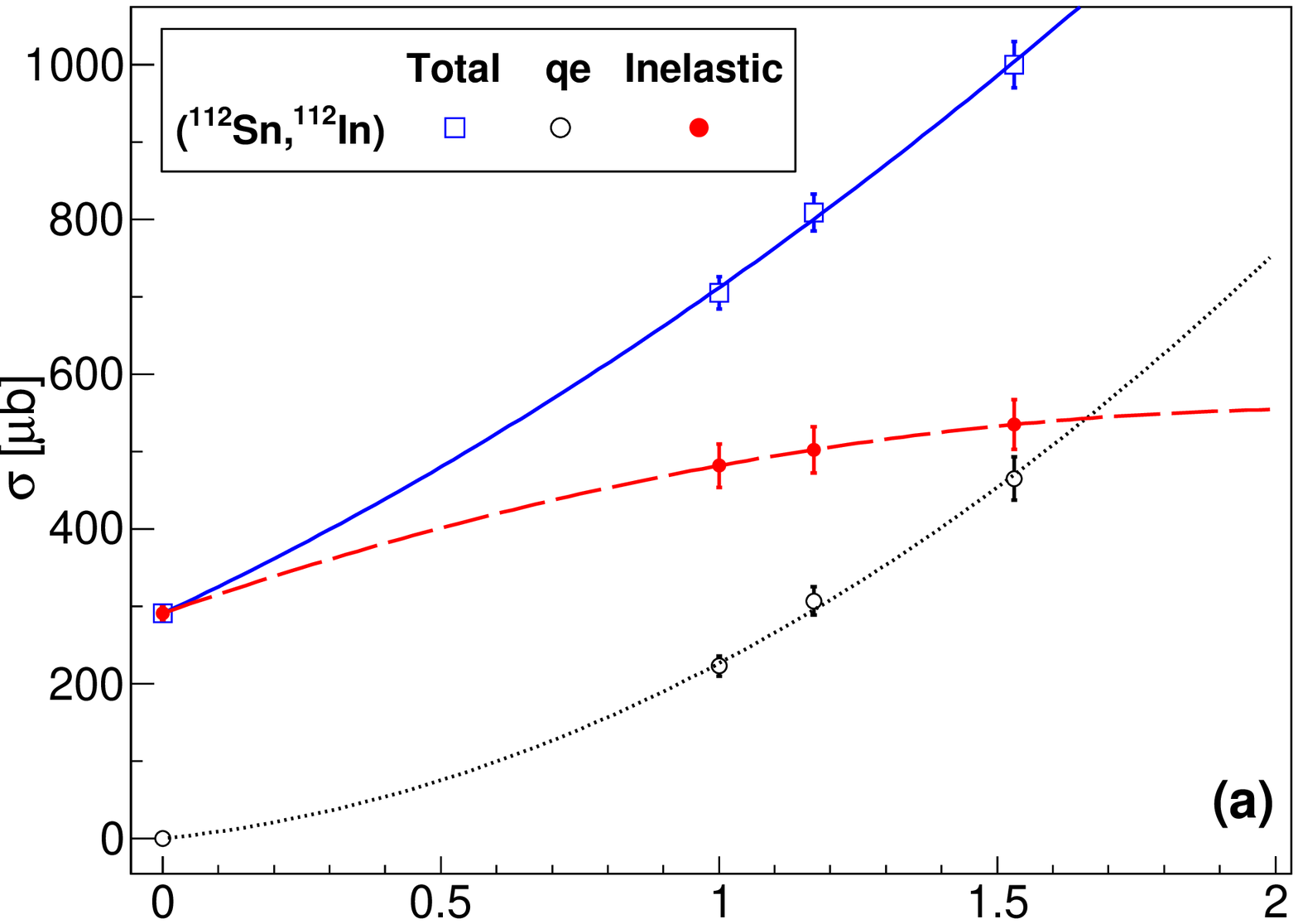}}
\centering
\subfigure{\label{fig:5b}\includegraphics[width=0.48\textwidth,keepaspectratio]{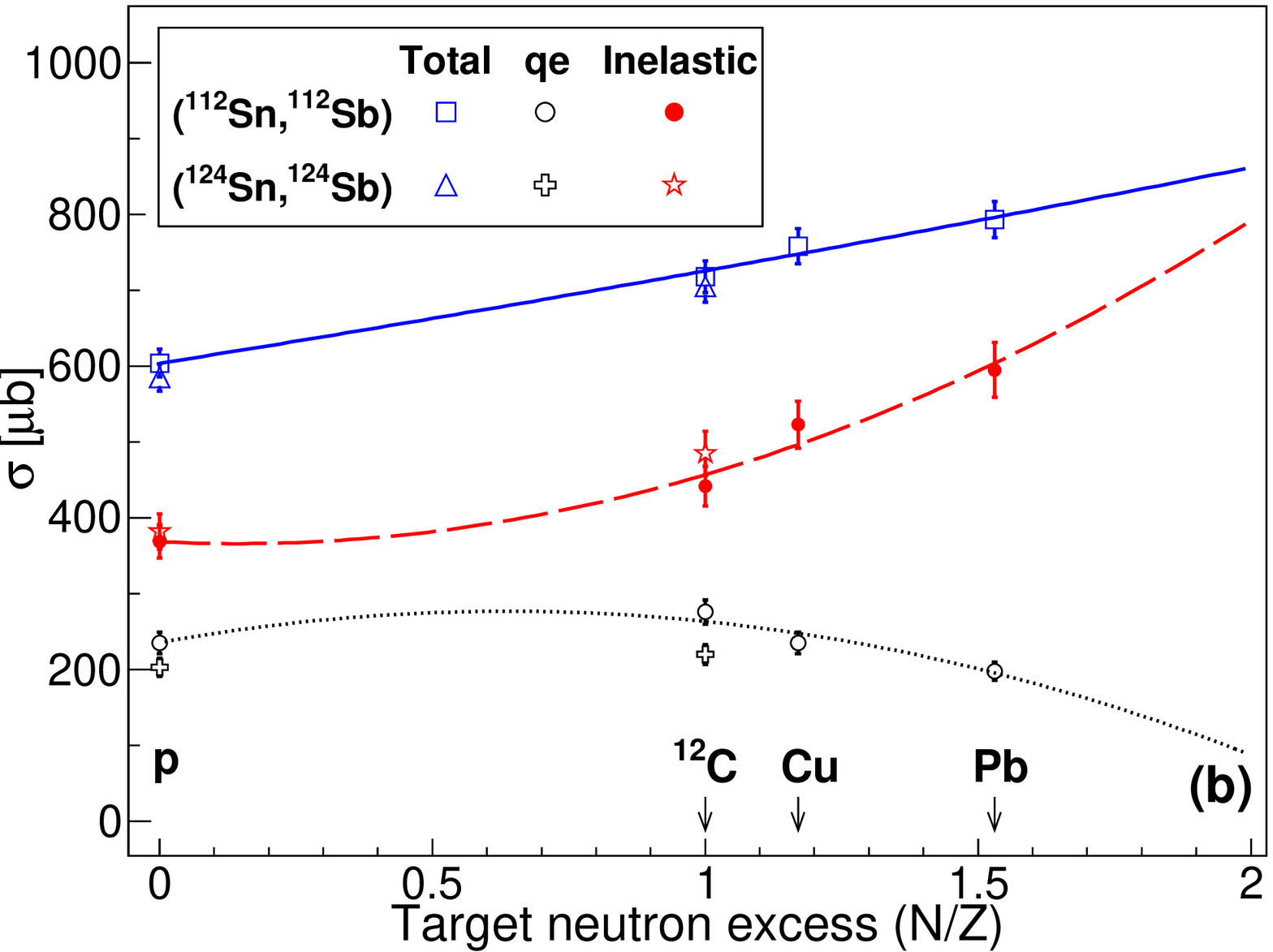}}
\centering
\caption{(Color online) Total, quasi-elastic (qe), and inelastic cross sections as a funtion of the neutron excess of target nuclei for the ($^{112}$Sn,$^{112}$In) (a) and ($^{112}$Sn,$^{112}$Sb) (b) reactions. We also compare these data to the cross sections obtained for the ($^{124}$Sn,$^{124}$Sb) reaction. Lines represent the best data fit. 
}
\label{fig:5}
\end{figure}

The dependence on the target neutron excess of the total isobaric charge-exchange cross section is illustrated in Figs.~\ref{fig:5a} and~\ref{fig:5b} for the ($^{112}$Sn,$^{112}$In) and ($^{112}$Sn,$^{112}$Sb) charge-exchange reactions, respectively. As expected, the cross section of the isobaric charge-exchange reactions increases with the size of the target nuclei, like it happens with the total interaction cross section~\cite{Ozawa2001}. In these two figures, we also display the cross sections for the quasi-elastic and inelastic components listed in Table~\ref{tab:1}, where we have fitted the quasi-elastic peak in the spectra to a generalized asymmetric Breit-Wigner function in order to extract its cross section, following the methodology proposed in Ref.~\cite{Seitz1970}. On the one hand, we can see that the quasi-elastic and inelastic cross sections for the ($^{112}$Sn,$^{112}$In) reaction increase with the target size, although this increase is steeper for the quasi-elastic peak. The best fit of the data also allows us to extrapolate the measurements to larger target neutron excess, where one can see that for very neutron-rich target nuclei, $N/Z$ larger than 1.7, the predominant contribution to the total cross section would be the quasi-elastic channel. On the other hand, for the ($^{112}$Sn,$^{112}$Sb) reaction we can observe that the quasi-elastic cross section decreases with the target neutron excess, nosediving to zero for the case of very neutron-rich target nuclei. This reduction could be attributed to the quenching of GT transitions as observed in other measurements performed with target nuclei with $A>12$~\cite{Gaarde1983}, which pointed out a quenching of around $50\%$. However, many collective isovector multipole modes are also contributing to both target and projectile excitations~\cite{PhysRevC.86.014304} and, therefore, it is very unlikely that this suppression can be only caused by the quenching of GT transitions. Additionally, the comparison to ($^{124}$Sn,$^{124}$Sb) in Fig.~\ref{fig:5b} also shows the relevance of the projectile neutron excess because the cross sections decrease systematically around $4\%$ (see Table~\ref{tab:1}), which could be attributed to the lower neutron separation energy in $^{124}$Sn~\cite{JL2017}. 

\section{Model calculations}
\label{sec:model}

The model employed to analyze the missing-energy spectra measured in the experiment was already introduced in our previous study of Ref.~\cite{Rodriguez2020}. We emphasize that charge-exchange reactions with nuclei require to account in parallel for $np^{-1}$ transitions in one ion (projectile or target) in combination with the complementary $pn^{-1}$ ones in the other nuclear system. A brief review of the model is given in the following.

The nuclear charge-exchange response functions of projectile and target for the quasi--elastic $np^{-1}$ and the  $pn^{-1}$ components, respectively, are obtained together with the $N^*N^{-1}$ spectral distributions in Random Phase Approximation (RPA) by solution of a system of coupled Dyson--equations. Since this approach is discussed in detail in Ref.~\cite{Lenske2018}, here we only discuss briefly the essential details. The Dyson--equations are solved in infinite asymmetric matter, accounting for blocking effects by using proton and neutron Fermi--spheres of the proper particle numbers. The non--elastic resonance components cover particle states given by the $\Delta_{33}(1232)$ resonance and the Roper $P_{11}(1440)$ resonance. The nucleonic RPA polarization propagators are obtained by mean--field and Landau--Migdal interactions derived consistently from the Giessen Energy Density Functional (GiEDF) \cite{Lenske:2019ubp,Adamian:2021gnm}, where interactions are extracted from a Brueckner G--matrix and supplemented by three--body interactions of the Urbana--type. Interactions between the $N^*N^{-1}$ configurations are described by $\pi$ and $\rho$--meson exchange. Short range nuclear effects are also taken into account by Landau parameters $g'_{N,N^*}$. All interactions include anti-symmetrization. The calculations also include dissipation effects through polarization self--energies for nucleons and resonances. As discussed in Ref.~\cite{Lenske2018}, the $N^*$ in--medium self--energies are corrected for Pauli--blocking effects in matter by using the Oset--Salcedo method~\cite{Oset:1987re}. The set of coupled RPA Dyson--equations is solved first in infinite nuclear matter. The resulting response functions, depending on the local proton and neutron densities $\rho_{p,n}$, are integrated over the nuclear volumes by replacing $\rho_{p,n}\mapsto \rho^{(HFB)}_{p,n}(r)$. In summary, response functions are obtained in local density approximation accounting for in--medium interactions, anti-symmetrization, and Pauli-blocking effects at all levels.

\begin{table}[t!]
\caption{Elementary processes included in the description of the isobaric charge-exchange reactions p($^{112,124}$Sn,$^{112,124}$Sb)X and $^{A}$Z($^{112,124}$Sn,$^{112,124}$Sb)X.
}
\label{tab:elp1}
\begin{center}
\small
\begin{tabular}{cc}
\hline
\hline
\\
\multicolumn{2}{c}{p($^{112,124}$Sn,$^{112,124}$Sb)X } \\ \\ \hline
\\
quasi-elastic &{$p(n,p)n$}
\\
\\
\hline
\\
\multicolumn{1}{c}{inelastic (target)} &  \multicolumn{1}{c}{inelastic (projectile)} \\ \\
\hline
\\
$p(n,p)\Delta^{0}\rightarrow p(n,p)n\pi^0$  & $p(n,\Delta^0)p\rightarrow p(n,p\pi^-)p$ \\
$p(n,p)\Delta^{0}\rightarrow p(n,p)p\pi^-$  & $p(n,\Delta^+)n\rightarrow p(n,p\pi^0)n$ \\
$p(n,p)n\pi^0$ (s-wave) & $p(n,p\pi^-)p$ (s-wave)  \\
$p(n,p)p\pi^-$ (s-wave) & $p(n,p\pi^0)n$ (s-wave)  \\
\\
\hline
\hline
\\
\multicolumn{2}{c}{$^{A}$Z($^{112,124}$Sn,$^{112,124}$Sb)X} \\ \\
\hline
\\
quasi-elastic &{$p(n,p)n$}
\\
\\
\hline
\\
\multicolumn{1}{c}{inelastic (target)} &  \multicolumn{1}{c}{inelastic (projectile)} \\
\\
\hline
\\
$p(n,p)\Delta^{0}\rightarrow p(n,p)n\pi^0$  & $p(n,\Delta^0)p\rightarrow p(n,p\pi^-)p$ \\
$p(n,p)\Delta^{0}\rightarrow p(n,p)p\pi^-$  & $p(n,\Delta^+)n\rightarrow p(n,p\pi^0)n$ \\
$p(n,p)n\pi^0$ (s-wave) & $p(n,p\pi^-)p$ (s-wave)  \\
$p(n,p)p\pi^-$ (s-wave) & $p(n,p\pi^0)n$ (s-wave)  \\
\\
$n(n,p)\Delta^-\rightarrow n(n,p)n\pi^-$  & $n(n,\Delta^0)n\rightarrow n(n,p\pi^-)n$ \\
$n(n,p)n\pi^-$ (s-wave) & $n(n,p\pi^-)n$ (s-wave)  \\
\\
\hline
\hline

\end{tabular}
\end{center}
\end{table}

\begin{table}[t!]
\caption{Elementary processes included in the description of the isobaric charge-exchange reactions p($^{112,124}$Sn,$^{112,124}$In)X and $^{A}$Z($^{112,124}$Sn,$^{112,124}$In)X.
}
\label{tab:elp2}
\begin{center}
\small
\begin{tabular}{cc}
\hline
\hline
\\
\multicolumn{2}{c}{p($^{112,124}$Sn,$^{112,124}$In)X } \\ \\ \hline
\multicolumn{1}{c}{inelastic (target)} &  \multicolumn{1}{c}{inelastic (projectile)} \\ \hline
\\
$p(p,n)\Delta^{++}\rightarrow p(p,n)p\pi^+$  & $p(p,\Delta^+)p\rightarrow p(p,n\pi^+)p$ \\
$p(p,n)p\pi^+$ (s-wave) & $p(p,n\pi^+)p$ (s-wave)  \\
\\
\hline
\hline
\\
\multicolumn{2}{c}{$^{A}$Z($^{112,124}$Sn,$^{112,124}$In)X} \\ \\
\hline
\\
quasi-elastic &{$n(p,n)p$}
\\
\\
\hline
\\
\multicolumn{1}{c}{inelastic (target)} &  \multicolumn{1}{c}{inelastic (projectile)} \\
\\
\hline
\\
$p(p,n)\Delta^{++}\rightarrow p(p,n)p\pi^+$  & $p(p,\Delta^+)p\rightarrow p(p,n\pi^+)p$ \\
$p(p,n)p\pi^+$ (s-wave) & $p(p,n\pi^+)p$ (s-wave)  \\ \\
$n(p,n)\Delta^+\rightarrow n(p,n)n\pi^+$  & $n(p,\Delta^+)n\rightarrow n(p,n\pi^+)n$ \\
$n(p,n)\Delta^+\rightarrow n(p,n)p\pi^0$  & $n(p,\Delta^0)n\rightarrow n(p,n\pi^0)p$ \\
$n(p,n)n\pi^+$ (s-wave) & $n(p,n\pi^+)n$ (s-wave) \\
$n(p,n)p\pi^0$ (s-wave)  & $n(p,n\pi^0)p$ (s-wave) \\
\\
\hline
\hline

\end{tabular}
\end{center}
\end{table}

\begin{figure}[h!]
\centering
\subfigure{
\includegraphics[width=0.43\textwidth,keepaspectratio]{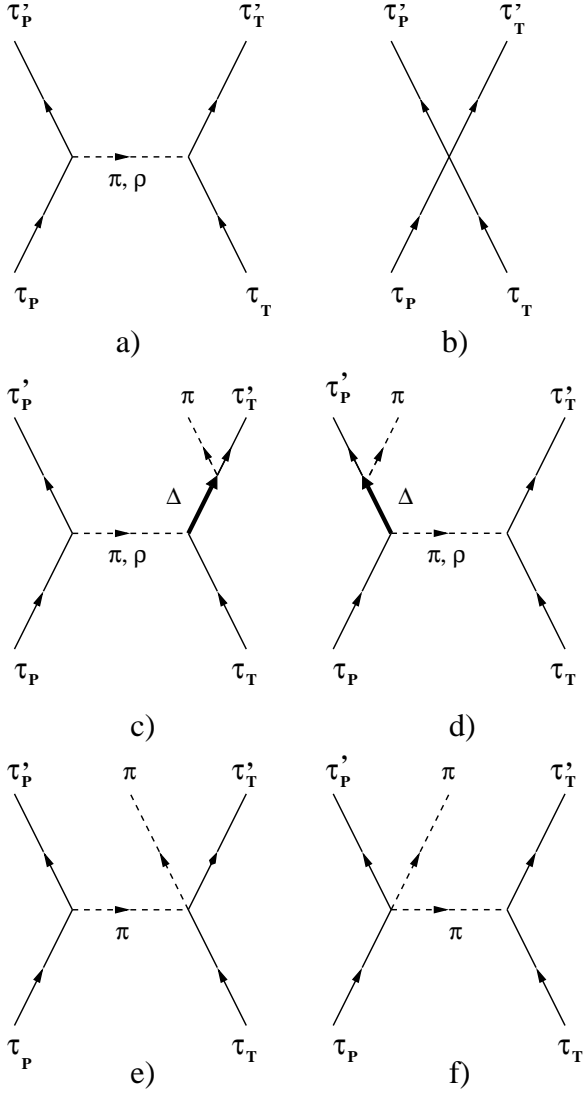}}
\caption{Direct contributions to the quasi-elastic (diagrams a and b), and inelastic (diagrams from c to f) $(n,p)$ and $(p,n)$ elementary processes. The emitted pion in the final state is assumed to come either from the decay $\Delta\rightarrow N\pi$ (diagrams c and d) or to be produced in s-wave without the excitation of any resonance (diagrams e and f). Exchange contributions are obtained by interchanging the lines of the incoming neutron or proton of the projectile and the nucleon $\tau$ of the target.
}
\label{fig:6}
\end{figure}

The reaction model takes into account the direct and exchange contributions to the missing-energy spectrum from quasi-elastic and inelastic $(n,p)$ and $(p,n)$ elementary processes, which are described in terms of the exchange of virtual $\pi$- and $\rho$-mesons between the interacting nucleons. The list of elementary processes included in the description of the $(n,p)$- and $(p,n)$-type charge exchange reactions is given in Tables~\ref{tab:elp1} and~\ref{tab:elp2}, respectively. The Feynman diagrams illustrating the direct contributions to these elementary processes are shown in Fig.~\ref{fig:6}. The exchange contributions can be simply obtained by interchanging the lines of the incoming nucleon $\tau_P$ of the projectile and the nucleon $\tau_T$ of the target. Note that the incoming projectile nucleon $\tau_P$ can only be a proton in the case of the p($^{112,124}$Sn,$^{112,124}$In)X reactions but it can be either a neutron or a proton in all other reactions considered. Diagram (a) gives the direct contribution to the quasi-elastic $(n,p)$ and $(p,n)$ elementary processes whereas diagram (b) takes into account the effect of short-range correlations by means of the Landau--Migdal parameter $g'$ taken here to be $0.7$. The excitation of the $\Delta$ resonance and its subsequent decay into a nucleon and a pion is considered both in the target (diagram c) and in the projectile (diagram d). Note that the label $\Delta$ in diagrams (c) and (d) indicates schematically the isospin states $\Delta^{++}, \Delta^+, \Delta^0$ or $\Delta^-$ depending on the particular elementary process (see Tables~\ref{tab:elp1} and~\ref{tab:elp2}). We note that contributions from the excitation of other resonances, such as for instance the Roper $P11(1440)$, have been ignored in the present model. Diagrams (e) and (f) show the contributions where the emitted pion is produced in s-wave. The basic ingredients of the model are the $NN\pi$, $NN\rho$, $N\Delta\pi$ and $N\Delta\rho$ vertices~\cite{Ericson87,Fernandez92} and the $\pi N\rightarrow \pi N$ s-wave amplitude \cite{Fernandez92,Oset89}. The inclusive cross section for the reaction $A(a,b)B$ describing the differential missing-energy spectrum of the outgoing projectile-like ion is given by the sum of a quasi-elastic (qe) and an inelastic (in) contribution
\begin{equation}
 \frac{d\sigma}{dE_{b}} = \frac{d\sigma}{dE_{b}} \Big |_{qe}+ \frac{d\sigma}{dE_{b}} \Big |_{in} \ .
\end{equation}
The quasi-elastic contribution is obtained as
\begin{equation}
 \frac{d\sigma}{dE_{b}} \Big |_{qe}=
\langle N_{\tau_P\tau_T}\rangle |\mathcal{M}_{qe}|^2
\label{eq:dxsqe}
\end{equation}
being $\langle N_{\tau_P\tau_T}\rangle$ (see Eq.~(\ref{eq:av}) below) the average number of neutron-proton ($\tau_P=n,\tau_T=p$) and proton-neutron ($\tau_P=p,\tau_T=n$) elementary processes contributing, respectively, to the $(n,p)$- and the $(n,p)$-type reactions, and
\begin{eqnarray}
&&\left|\mathcal{M}_{qe}\right|^2=\frac{2|{\vec p}_b|}{(2\pi)^2}\frac{m^4}{\lambda^{1/2}(s,m^2,m^2)}   \nonumber \\
&&\times
\int  \frac{d\vec q }{(2\pi)^3}\frac{d\Omega_{b}}{E_{B}}
R^{(A)}_N(\sqrt{s}-E_b,\vec{q})\langle \left|M_{qe}(\vec {q})\right|^2\rangle \ , \nonumber \\
\label{eq:dxsqe}
\end{eqnarray}
where $s$ is the total energy in the center-of-mass frame, $m$ is the nucleon mass, $\langle \cdot \rangle$ indicates the average and sum over the initial and final spins, $R^{(A)}_N(\omega,\vec {q})$ is the spectral distribution of nucleon hole-nucleon $N'N^{-1}$ target transitions (see Refs.\  \cite{Lenske2018,Lenske18b,Lenske18c,Lenske2019} for detailed discussions and explicit expressions), $M_{qe}$ is the scattering amplitude of the elementary process $p+n\rightarrow n+p$ and
$\lambda(a,b,c)=a^2+b^2+c^2-2ab-2ac-2bc$ is the so-called K\"{a}llen function.

The inelastic contribution is given by
\begin{widetext}
\begin{eqnarray}
 \frac{d\sigma}{dE_{b}} \Big |_{in}=\left\{\begin{array}{lll}
\langle N_{nn}\rangle |\mathcal{M}_{in}^{(nn\rightarrow pn\pi^-)}|^2
+\langle N_{np}\rangle |\mathcal{M}_{in}^{(np\rightarrow pp\pi^-)}|^2
+\langle N_{np}\rangle |\mathcal{M}_{in}^{(np\rightarrow pn\pi^0)}|^2
\, , & & \mbox{for $(n,p)$-type reactions} \\ \\
\langle N_{pp}\rangle |\mathcal{M}_{in}^{(pp\rightarrow np\pi^+)}|^2
+\langle N_{pn}\rangle |\mathcal{M}_{in}^{(pn\rightarrow nn\pi^+)}|^2
+\langle N_{pn}\rangle |\mathcal{M}_{in}^{(pn\rightarrow np\pi^0)}|^2
\, , & & \mbox{for $(p,n)$-type reactions}
\end{array}
\right. 
\label{eq:dxsin1}
\end{eqnarray}
\end{widetext}
with $\langle N_{nn}\rangle$ and $\langle N_{pp}\rangle$, similarly to  $\langle N_{np}\rangle$ and $\langle N_{pn}\rangle$, being the average number of  neutron-neutron and proton-proton elementary processes contributing to the $(n,p)$ and $(p,n)$ reactions, defined also through Eq.~(\ref{eq:av}) and
\begin{eqnarray}
&&\left|\mathcal{M}_{in}^{(\tau_P\tau_T\rightarrow \tau'_P\tau'_T\pi)}\right|^2= \frac{1}{S}\frac{|{\vec p}_b|}{(2\pi)^5}\frac{m^4}{\lambda^{1/2}(s,m^2,m^2)}  \nonumber \\
&&\times
\int\frac{d\vec q }{(2\pi)^3} \frac{d\vec p_\pi d\Omega_{b}}{E_{B}E_\pi}   \nonumber \\
&&\times R^{(A)}_{\Delta}(\sqrt{s}-E_b-E_\pi,\vec{q}-\vec{p}_\pi) \nonumber \\
&&\times\langle \left|M_{in}^{ (\tau_P\tau_T\rightarrow \tau'_P\tau'_T\pi )}(\vec{q}-\vec{p}_\pi) \right|^2\rangle \ . \nonumber \\
\label{eq:dxsin}
\end{eqnarray}

Here $R^{(A)}_\Delta(\omega,\vec{q})$ is the spectral distribution of $\Delta$ hole-nucleon $\Delta N^{-1}$ target transitions, $M_{in}^{(\tau_P\tau_T\rightarrow \tau'_P\tau'_P\pi)}$ is the scattering amplitude of the elementary process $\tau_P+\tau_T \rightarrow \tau'_P+\tau'_P+\pi$ and $S$ is a symmetry factor given by
\begin{equation}
S=\prod_{l}k_l! \nonumber
\end{equation}
for $k_l$ identical particles of species $l$ in the final state. In our case $S$ can be 1 or 2 depending on the particular reaction channel (see Tables~\ref{tab:elp1} and~\ref{tab:elp2}).

The average number of elementary processes contributing to the $(n,p)$- and $(p,n)$-type reactions is calculated as
\begin{equation}
\langle N_{\tau_P\tau_T}\rangle=\langle N_{\tau_P}\rangle\times\langle N_{\tau_T}\rangle \ , \,\,\,\, \tau_P=n,p \ , \,\, \tau_T=n,p
\label{eq:av}
\end{equation}
where $\langle N_{\tau_P}\rangle$ and $\langle N_{\tau_T}\rangle$ are, respectively, the average number of neutrons or protons in the projectile and in the target participating in the reaction. These numbers are calculated as
\begin{eqnarray}
\langle N_{n_P}\rangle&=&(A_P-Z_P)\frac{\sigma_T}{\sigma_{PT}}  \nonumber  \\
\langle N_{p_P}\rangle&=&Z_P\frac{\sigma_T}{\sigma_{PT}}  \nonumber  \\
\langle N_{n_T}\rangle&=&(A_T-Z_T)\frac{\sigma_P}{\sigma_{PT}} \nonumber \\
\langle N_{p_T}\rangle&=&Z_T\frac{\sigma_P}{\sigma_{PT}}  \ ,\nonumber
\end{eqnarray}
where $A_P$ $(Z_P)$  and $A_T$ $(Z_T)$ are the mass (atomic) numbers of the projectile and target nuclei, respectively, and  $\sigma_{P(T)}$ and $\sigma_{PT}$ the total nucleon-nucleus and nucleus-nucleus cross section
determined by using the Glauber model
\begin{equation}
\sigma_{P(T)}=\int d\vec b \Big(1-(1-T_{P(T)}(b)\sigma_{NN})^{A_{P(T)}} \Big)\nonumber
\end{equation}
\begin{equation}
\sigma_{PT}=\int d\vec b \Big(1-(1-T_{PT}(b)\sigma_{NN})^{A_{P}A_{T}} \Big)\nonumber
\end{equation}
with $\sigma_{NN}$ the nucleon-nucleon cross section, for which we take here a value of 40 mb, and
\begin{equation}
T_{P(T)}(b)=\int dz \rho_{P(T)}(b,z)\nonumber
\end{equation}
\begin{equation}
T_{PT}(b)=\int d\vec s T_P(|\vec s -\vec b|)T_T(s) \ ,\nonumber
\end{equation}
with the projectile and target densities $\rho_P$ and $\rho_T$ obtained from a relativistic mean field model calculation using the FSU model~\cite{Todd05}.

\section{Discussion}
\label{sec:disc}
 
\begin{figure*}
\centering
\subfigure{
\includegraphics[width=0.99\textwidth,keepaspectratio]{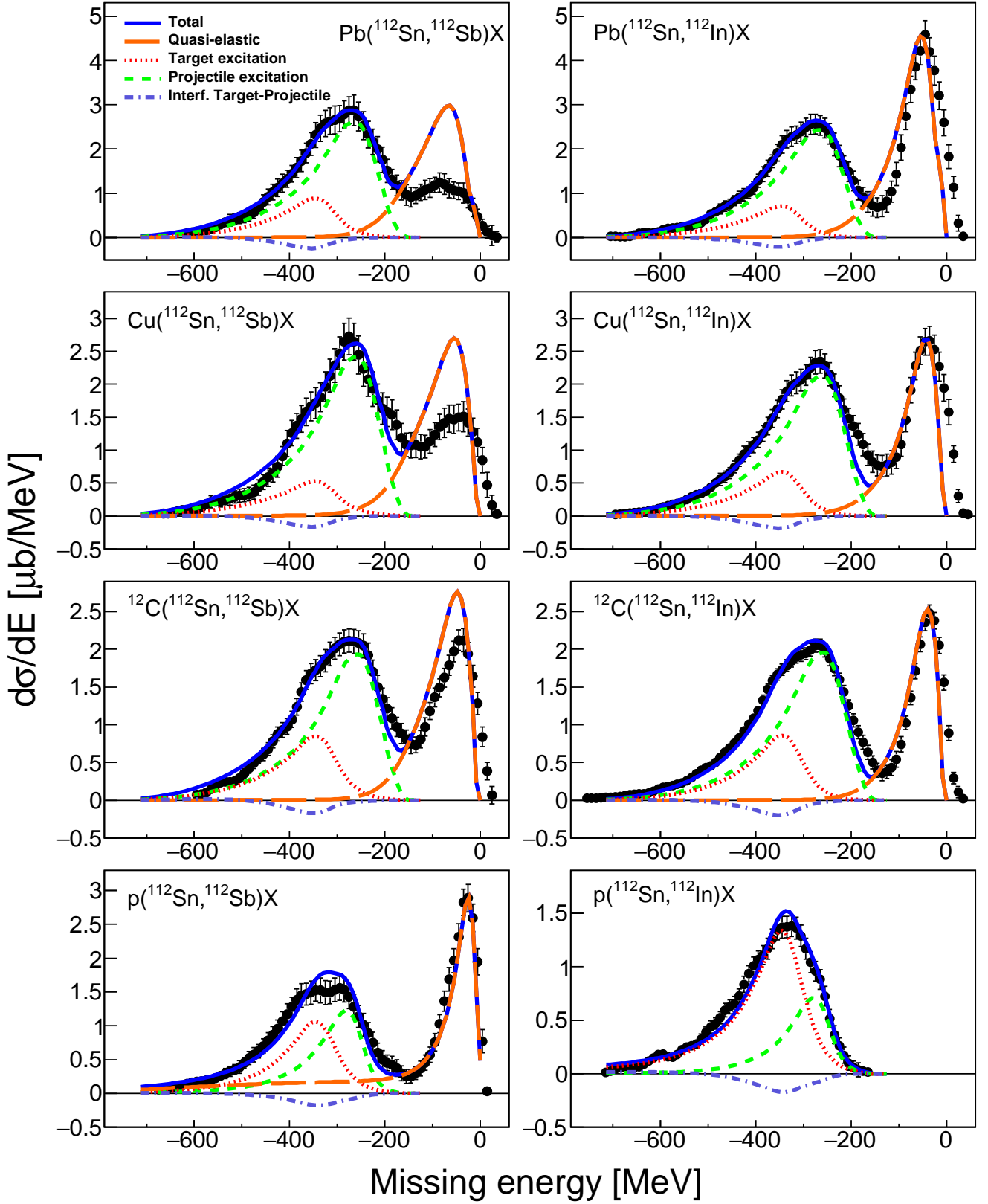}}
\caption{(Color online) Missing-energy spectra obtained from the Pb, Cu, $^{12}$C and proton targets for the single isobaric charge-exchange reactions ($^{112}$Sn,$^{112}$Sb) and ($^{112}$Sn,$^{112}$In) compared to the theoretical calculations described in Ref.~\cite{Rodriguez2020}. See the legend displayed in the upper left panel for explanation.
}
\label{fig:7}
\end{figure*}

The comparison of the experimental data for the single isobaric charge-exchange reactions ($^{112}$Sn,$^{112}$Sb) and ($^{112}$Sn,$^{112}$In) with the model calculations is shown in Fig.~\ref{fig:7}. The separate contributions from the quasi-elastic and inelastic (target and projectile excitation) processes are shown, as well as the interference between the target and projectile excitation processes. We should mention that in the calculation for the $(p,n)$-type reactions these contributions have been rescaled separately to compare with the data. For the $(n,p)$-type ones, however, a common rescaling factor for these contributions has been considered. The reason for this is to check whether the model can explain the quenching of the quasi-elastic peak observed in the data of the ($^{112}$Sn,$^{112}$Sb) reactions without adjusting it {\it ad hoc}. We note that although the shape of the experimental spectrum is in general reasonably well reproduced by the model, the quenching of the quasi-elastic peak cannot be described. Since many multipoles are contributing to both target and projectile excitations, it is very unlikely that the suppression is caused by the quenching of GT transitions as observed in Ref.~\cite{Gaarde1983}. From the theoretical point of view this effect is not fully understood, in particular because the response functions account correctly for the differences in the available configuration spaces of the $pn^{-1}$ and $np^{-1}$ transitions for the projectile and target nuclei, respectively. In the following, we will see that the model allows us to reveal a large amount of information on the different reaction mechanism that lead to the inclusive spectra measured.

We consider first the spectra obtained with the proton target. The first observation is, as expected, the absence of the quasi-elastic peak at missing-energies close to zero in the case of the p($^{112}$Sn,$^{112}$In)X reaction whereas it is clearly present in the p($^{112}$Sn,$^{112}$Sb)X one. At lower missing energies the spectrum of both reactions can be explained as a result of the superposition of the four (see Table~\ref{tab:elp2}) and eight (see Table~\ref{tab:elp1}) contributing elementary processes, respectively. As it can be seen in Fig.~\ref{fig:7}, the spectrum of the p($^{112}$Sn,$^{112}$In)X reaction is clearly dominated by the excitation of the $\Delta^{++}$ in the target (see Table~\ref{tab:elp2}), although the contribution of the other three elementary processes, particularly the $\Delta^+$ excitation in the projectile, is necessary to reproduce its shape. This is not the case for the p($^{112}$Sn,$^{112}$Sb)X reaction where the contribution due to the excitation of the $\Delta^{0}$ in the target (see Table~\ref{tab:elp1}) and the excitation of the $\Delta^{+}$ and $\Delta^{0}$ in the projectile are similar as reflected by the double-peak structure seen in the data. Note, in addition, that the small shoulder observed in the p($^{112}$Sn,$^{112}$In)X reaction at the right of the peak can be understood mainly due to the contribution of the $\Delta^+$ excitation in the projectile. In both reactions, the contribution from the processes where the pion is emitted in a s-wave is small in comparison to those of the excitation of the $\Delta$ isobar. Moreover, one can see that the shape and the position of the resonance is different if it is excited in the target or in the projectile nucleus. This is because its invariant mass is different in both cases. When the $\Delta$-resonance is excited in the target, its invariant mass does not depend on the momentum of the emitted pion and, therefore, the scattering amplitude can be taken out of the integral in Eq.~(\ref{eq:dxsin}). In this case, the shape of the resonance appears almost symmetric and it has a peak at approximately -345~MeV. On the contrary, when the $\Delta$ is excited in the projectile, its invariant mass depends explicitly on the momentum of the emitted pion and, therefore, it is clear here that the scattering amplitude must also be integrated. Consequently, the resonance acquires an apparent asymmetric shape and its position is shifted to the right by $\sim$63 MeV. This is precisely the shift in the position of the $\Delta$ peak that seems to be observed between the measurements with the proton and heavier targets.

\begin{figure}[t!]
\centering
\subfigure{\label{fig:8a}\includegraphics[width=0.48\textwidth,keepaspectratio]{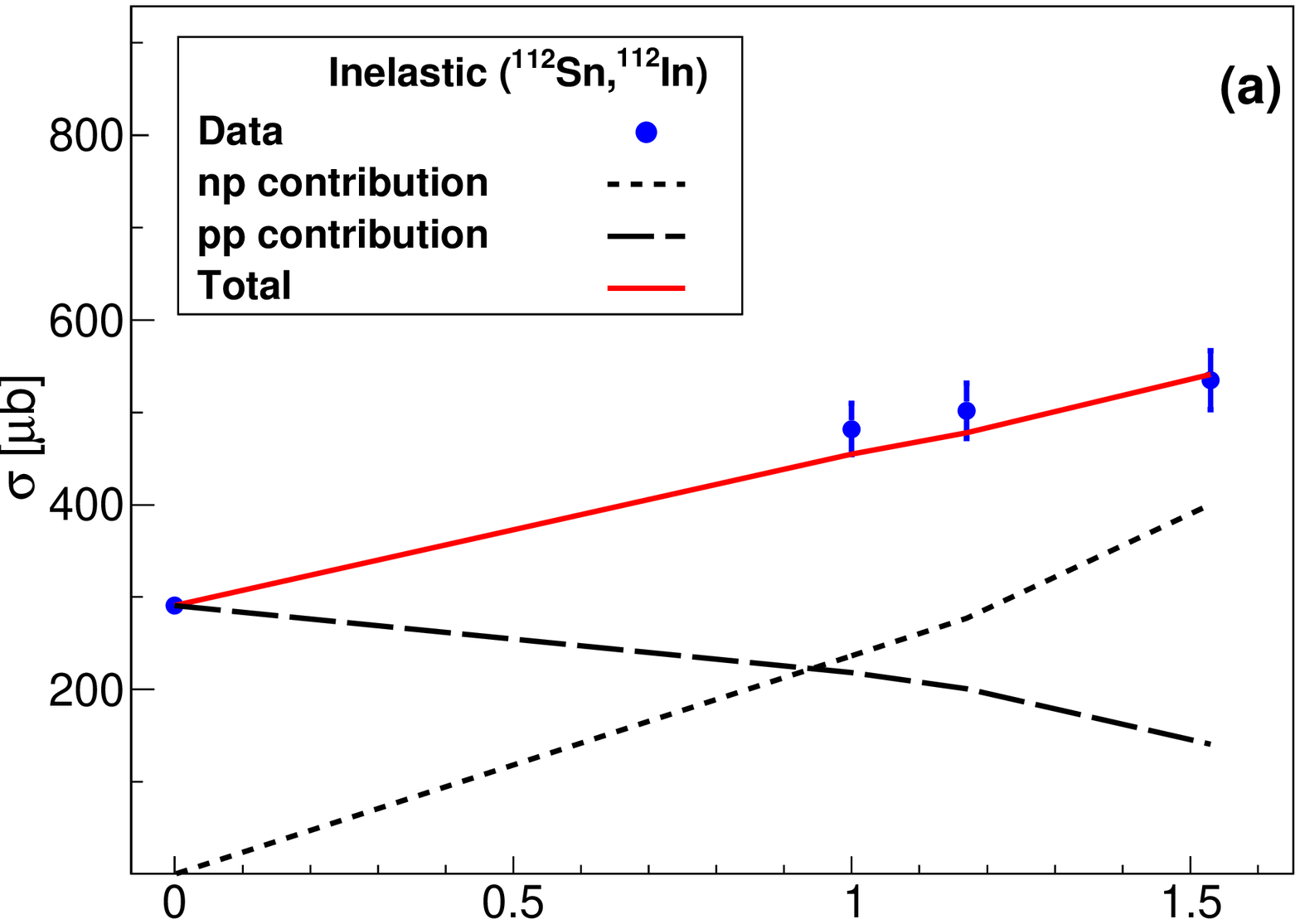}}
\centering
\subfigure{\label{fig:8b}\includegraphics[width=0.48\textwidth,keepaspectratio]{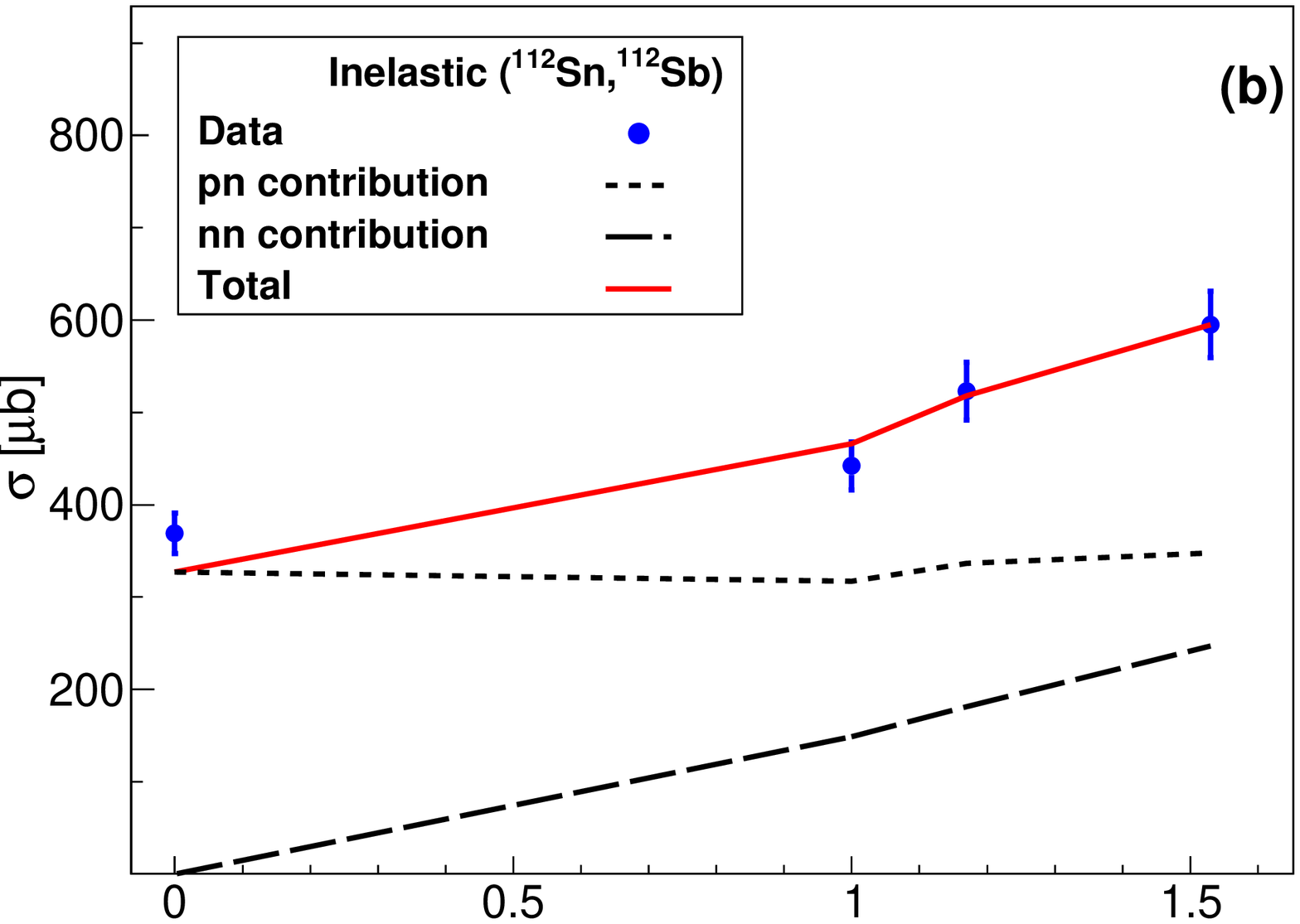}}
\centering
\subfigure{\label{fig:8c}\includegraphics[width=0.48\textwidth,keepaspectratio]{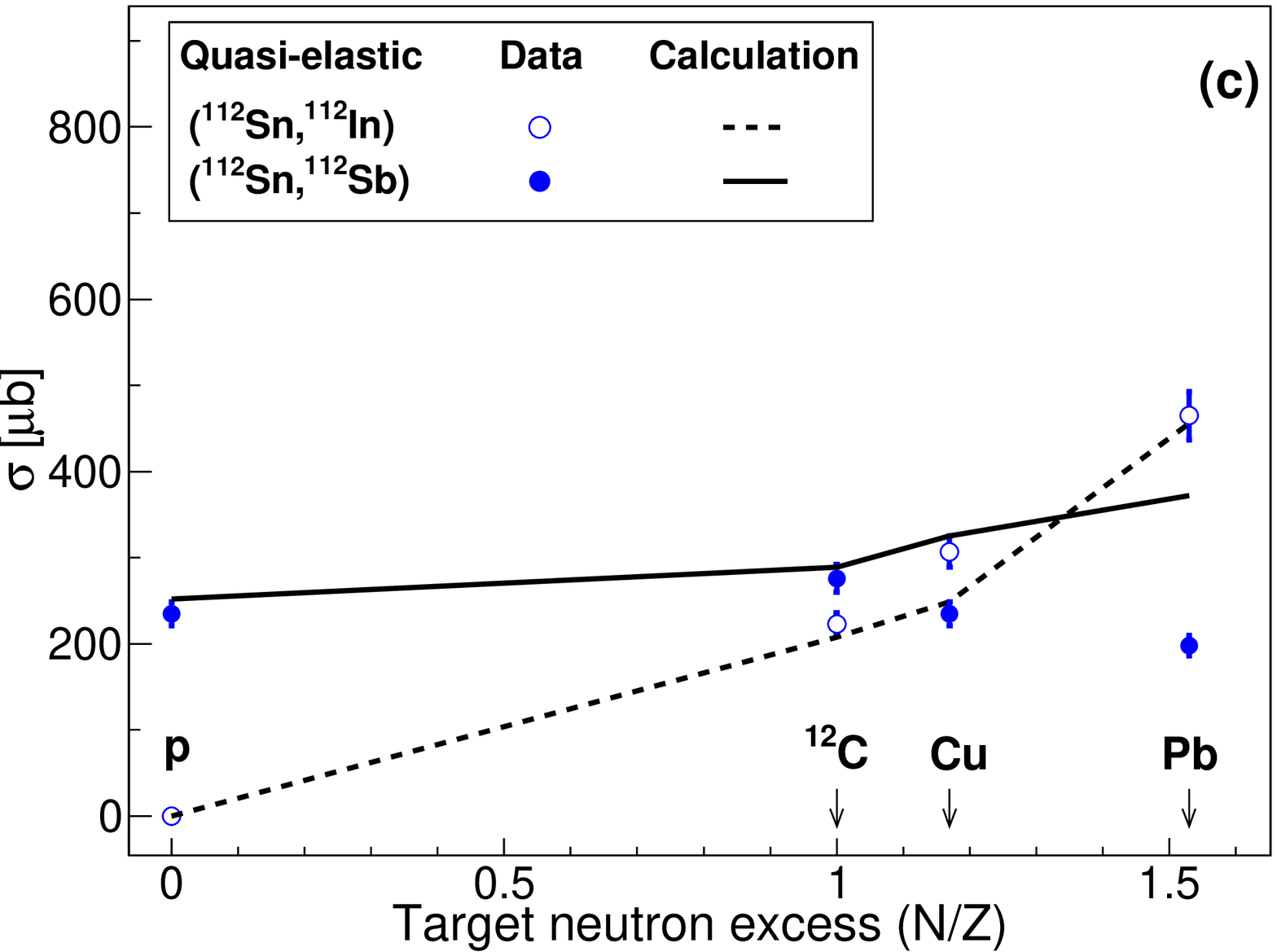}}
\centering
\caption{(Color online) Inelastic cross sections as a function of the neutron excess of target nuclei for the ($^{112}$Sn,$^{112}$In) (a) and ($^{112}$Sn,$^{112}$Sb) (b) reactions. The lines represent the contributions from np, pp, pn, and nn interactions predicted by our theoretical model (see Tables~\ref{tab:elp1} and~\ref{tab:elp2}). (c) Quasi-elastic cross sections for the ($^{112}$Sn,$^{112}$In) and ($^{112}$Sn,$^{112}$Sb) reactions compared to our calculations.
}
\label{fig:8}
\end{figure}

Let us finally analyze the results of the ($^{112}$Sn,$^{112}$Sb) and ($^{112}$Sn,$^{112}$In) reactions on heavier targets. We observe first, as mentioned before, that whereas the quasi-elastic peak is reasonably well reproduced by the model in the case of the $(^{112}$Sn,$^{112}$In) reactions, its quenching in the ($^{112}$Sn,$^{112}$Sb) ones cannot be described, likely due to the uncertainties in the momentum distributions of the nucleons involved in the collision. These uncertainties would be attributed to the re-scattering of the interacting nucleons with the rest of nucleons (final state interactions)~\cite{Diaz2020}, or short range nucleon-nucleon correlations~\cite{CLAS}. The strength in the lower missing-energy region of the spectrum is once more the result of the superposition of elementary processes where the excitation ($\Delta$ resonance or pion production in s-wave) occurs either in the target or in the projectile. Note that in this case there are additional elementary processes (see Tables~\ref{tab:elp1} and~\ref{tab:elp2}) contributing that were absent in the case of the reactions with the proton target, explaining the larger magnitude of the spectrum, apart from the trivial scaling by the increased number of participating target nucleons. In addition, there is a change of the relative magnitude between the excitation in target and projectile, with the latter one dominating now in both ($^{112}$Sn,$^{112}$Sb) and ($^{112}$Sn,$^{112}$In) reactions. The reason is that in a complex target nucleus only part of the energy transferred goes into the excitation of the $\Delta$ resonance or the direct s-wave production of the pion. Besides the excitations of $N^{-1}\Delta$ configurations, another part of the energy loss is employed in reactions such as, for instance, the knockout of nucleons from the target or its fragmentation. Consequently, this contribution is reduced with respect to that of the projectile excitation and is now observed simply as a shoulder at $\sim$ 63 MeV to the left of the dominant peak corresponding to the excitation of the $\Delta$ in the projectile. This seems to indicate that the apparent shift in the position of the $\Delta$-resonance peak, observed in targets heavier than the proton, can be simply interpreted as a change in the relative magnitude between the contribution of the excitation of the resonance in the target and in the projectile nuclei. Of course, only future exclusive measurements, allowing the separate identification of the resonance excitation in target and projectile nuclei, will be able to confirm this conclusion. We would like to emphasize that a similar interpretation was already pointed out by Oset, Shiino, and Toki~\cite{Oset89} in their analysis of the ($^3$He,t) reaction on proton, deuteron, and carbon targets. 

Finally, we also compare the measured inelastic cross sections of the ($^{112}$Sn,$^{112}$In) and ($^{112}$Sn,$^{112}$Sb) reactions to our theoretical calculations in Figs.~\ref{fig:8a} and~\ref{fig:8b}, respectively, showing the np, nn, and pp contributions from the reactions listed in Tables~\ref{tab:elp1} and~\ref{tab:elp2}. This allows us to investigate the neutron abundance effects in the $\Delta$-resonance excitation. For the reaction ($^{112}$Sn,$^{112}$In), we can see that the pn contribution to the $\Delta$ excitation increases linearly with the target neutron excess while the pp contribution decreases. This is expected since pn collisions are most likely when the target nuclei contain more neutrons at their nuclear surface. In the case of ($^{112}$Sn,$^{112}$Sb) reactions we observe that the np contribution is more or less constant with the target neutron excess while the nn collisions increase linearly. The same evolutions with the neutron excess are also observed for the np collisions involved in the quasi-elastic excitations, as shown in Fig.~\ref{fig:8c}. However, in the case of ($^{112}$Sn,$^{112}$Sb) only the inelastic cross sections are well described. The decreasing value of the measured quasi-elastic cross sections is not reproduced by the model calculations as mentioned afore.

\section{Conclusions and perspectives}
\label{sec:conclusions}

Single isobaric charge-exchange reactions induced by stable projectiles of $^{112,124}$Sn at energies of $1A$ GeV on different targets have been investigated at the GSI facility by using the magnetic spectrometer FRS. The experimental setup allowed us to measure the cross sections of these charge-exchange reactions with an accuracy of around $3\%$ and to determine simultaneously the missing-energy spectra of the corresponding ejectiles. Thanks to the FRS high-resolving power we can clearly identify in the missing-energy spectra the quasi-elastic and inelastic components, corresponding to the nuclear spin-isospin response of nucleon-hole and $\Delta$-resonance excitations, respectively. 

For both Sn projectiles, we observe an apparent energy shift for the $\Delta$-resonance peak of about (63$\pm$5) MeV when comparing the missing-energy spectra obtained from the measurements with the proton target to the other ones. This observation is consistent with the results obtained in the SATURNE experiments~\cite{Contardo1986}. However, the detailed analysis with our theoretical model indicates that this energy shift can be simply interpreted as a change in the relative magnitude between the $\Delta$-resonance excitation in target and projectile nuclei. We have also seen that for the $(p, n)$-type isobaric charge-exchange channel the quasi-elastic and inelastic cross sections increase with the target neutron excess, which can be explained by the increase of the interaction cross section with the size of target nuclei. However, for the $(n, p)$ channel we find a decrease of the quasi-elastic cross section with the target neutron excess that clearly indicates a quenching of this transition. At first glance, the quenching factor of about $50\%$ seems to indicate an effect similar to the widely observed suppression of GT-strength~\cite{Gaarde1983}. However, this explanation seems rather unlikely in this case because a multitude of multipolarities contributes to the reaction.

\begin{figure}[t!]
\centering
\subfigure{
\includegraphics[width=0.48\textwidth,keepaspectratio]{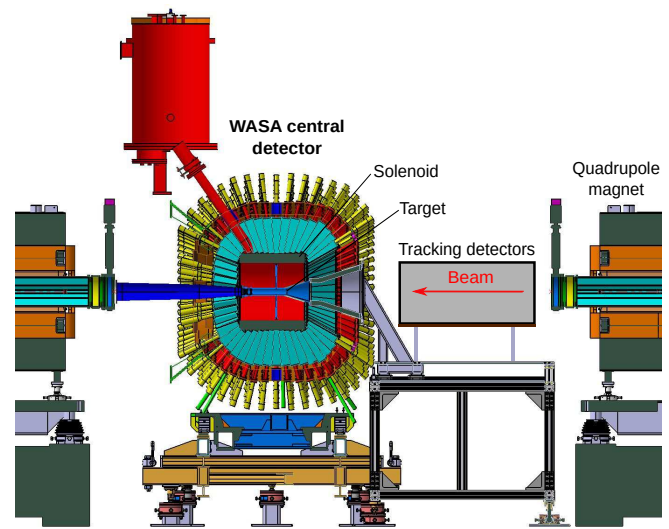}}
\caption{(Color online) Schematic illustration of the WASA calorimeter at the intermediate focal plane of Super-FRS. This setup consists of a mini-drift chamber made of straw tubes, a counter barrel made of plastic scintillators, a superconducting solenoid magnet, and an external iron yoke. Upstream of the WASA device we will install tracking detectors to identify the incoming projectiles.
}
\label{fig:9}
\end{figure}

Further investigations require the combination of the magnetic spectrometer with pion detection systems, like the WASA (Wide Angle Shower Apparatus) calorimeter designed originally to measure light ion collisions and $\eta$ rare decays at the CELSIUS storage ring in Uppsala (Sweden)~\cite{Bargholtz2008} and, more recently, at COSY in J\"{u}lich (Germany)~\cite{Cosy2013}. In 2019, this calorimeter was moved from J\"{u}lich to GSI and is now being installed at the intermediate focal plane of the FRS to study the properties of light hypernuclei and mesonic nuclei~\cite{YT2020}. In the future it will also be installed at the intermediate focal plane of the Super-FRS~\cite{Winfield2021} as shown in Fig.~\ref{fig:9}. This calorimeter would be used to measure the emitted pions from the baryonic resonance decays and non-resonant excitation channels in coincidence with the isobaric charge-exchange residues. For producing the reactions, the targets can be located inside the calorimeter. Surrounding the target we will have the WASA detectors and other passive elements whose details can be found in Ref.~\cite{Bargholtz2008}. The WASA detectors will provide us the momenta of the pions that combined with the momenta of the isobaric charge-exchange residues will give us the possibility of reconstructing the resonance invariant mass with resolutions of $\Delta M/M \sim 10^{-3}$.

Finally, these exclusive measurements tagging the pions will allow us to distinguish between target and projectile excitations unambiguously, giving access to the baryon invariant mass reconstruction and confirming whether the apparent shift in the position of the $\Delta$-resonance peak, observed in targets heavier than the proton, can indeed be interpreted as a change in the relative magnitude between the processes where the $\Delta$-resonance is excited in the target or in the projectile. Moreover, we will also study the in-medium modifications of the mass and width of the resonances, as well as to perfom missing mass spectroscopy of neutral decay channels involving neutron or $\pi^0$ emissions. This methodology was already proposed by Hennino and collaborators at the SATURNE facility where they observed a clear target mass dependence of the invariant mass~\cite{Hennino1992}, resulting in a $\Delta$-mass reduction of $\approx$25 MeV with respect to the elementary pn process. Additionally, this research will also offer the possibility of studying the in-medium decay properties of other resonances, like the Roper P11(1440)~\cite{Roper1964,Morsch1992,Alvarez1998}. In conclusion, these kinds of experiments with radioactive ion beams represent unique opportunities to understand in detail the formation of baryonic resonances in the nuclear medium and will give us for the first time access to systematically study their dynamics and properties with the nuclear isospin, covering large ranges in neutron-proton asymmetry.

\begin{acknowledgments}
The authors are grateful to the GSI accelerator staff for providing intense and stable beams of $^{112,124}$Sn. This work was partially supported by the Spanish Ministry for Science and Innovation under grants RTI2018-101578-B-C21 and PGC2018-099746-B-C22, and by the Regional Government of Galicia under the program “Grupos de Referencia Competitiva” ED431C-2021-38. This work has received financial support from Xunta de Galicia (Centro singular de investigaci\'on de Galicia accreditation 2019-2022). I.V. thanks the support from the \textit{COST Action CA16214}. H.L. acknowledges support by DFG, Grant Le439$/$16-2. J.L.R.S. thanks the support from Xunta de Galicia under the program of postdoctoral fellowships ED481B-2017-002 and ED481D-2021-018.
\end{acknowledgments}


\end{document}